\documentclass[aps,pra,twocolumn,groupedaddress,showpacs]{revtex4-1}
\usepackage{graphicx}
\usepackage{amsmath}
\usepackage{amsfonts}
\usepackage{amssymb}
\usepackage{ulem}

\usepackage{xcolor}
\usepackage{float}

\def\ket#1{{\left\vert#1\right\rangle}}
\def\abs#1{{\left|#1\right|}}

\def\tFWHM{\tau_\text{FWHM}}
\def\tctrl{\tau^\text{ctrl}_\text{FWHM}}
\def\Deltau{\Delta\tau^\text{ctrl}}

\begin{document}

\title{Optimization of Broadband $\Lambda$-type Quantum Memory Using Gaussian Pulses}
\author{Kai Shinbrough$^{1,2}$}
\email{kais@illinois.edu}
\author{Benjamin D. Hunt$^{1,2}$}
\author{Virginia O. Lorenz$^{1,2}$}

\address{$^1$Department of Physics, University of Illinois at Urbana-Champaign, 1110 West Green Street, Urbana, IL 61801, USA\\
$^2$Illinois Quantum Information Science and Technology (IQUIST) Center, University of Illinois at Urbana-Champaign, 1101 West Springfield Avenue, Urbana, IL 61801, USA}

\date{\today}

\begin{abstract}
    Optical quantum memory---the ability to store photonic quantum states and retrieve them on demand---is an essential resource for emerging quantum technologies and photonic quantum information protocols. Simultaneously achieving high efficiency and high-speed, broadband operation is an important task necessary for enabling these applications. We investigate the optimization of a large class of optical quantum memories based on resonant and near-resonant interaction with ensembles of $\Lambda$-type level systems with the restriction that the temporal envelope of all optical fields must be Gaussian, which reduces experimental complexity. Through this optimization we demonstrate an experimentally simple path to saturation of the protocol-independent storage efficiency bound that is valid for a wide range of memory bandwidths, including those that are broadband and high-speed. Examining the resulting optimal Gaussian control field parameters, we find a continuous transformation between three physically distinct resonant quantum memory protocols. We compare this Gaussian optimization scheme with standard shape-based optimization.
\end{abstract}

\maketitle

\section{Introduction}

Efficient photonic quantum state generation and synchronization \cite{Nunn13,Kaneda19}, metropolitan-scale quantum networking and entanglement distribution \cite{Sangouard11,DLCZ}, and linear-optical quantum computing \cite{KLM} all rely on efficient optical quantum memory. In order for these emerging applications to operate at high speed they must be compatible with broadband photonic quantum states \cite{Reim10, Simon10, BroadbandEIT}, ideally with minimal experimental complexity. In quantum memories based on atomic ensembles, a significant body of theoretical \cite{Novikova07, Gorshkov07, Gorshkov07_2, Gorshkov08, Nakao17, NunnThesis} and experimental \cite{Novikova08,Phillips08,Guo19} work has been dedicated to improving quantum memory efficiency by temporal shaping of the optical signal field to be stored or the control field used to mediate the interaction. However, these techniques have largely only been applied for signal bandwidths smaller than the linewidths of the excited states participating in the memory interaction, in part due to the technological complexity of shaping intense broadband fields. In effect, ensemble quantum memories to date have been limited to efficient narrowband operation \cite{Hosseini11,Chen13,Cho16,Hsiao18,Wang19} or inefficient broadband operation \cite{Reim10,England13,Bustard13,England15, Michelberger15,Thomas17,Saglamyurek11,Fisher17,Kaczmarek18,BinBaMemory,Ding15}, with only a few notable exceptions \cite{ATS,ATS2,ATS3,ATS4,BroadbandEIT}.

\begin{figure}[t]
	\centering
	\includegraphics[width=\columnwidth]{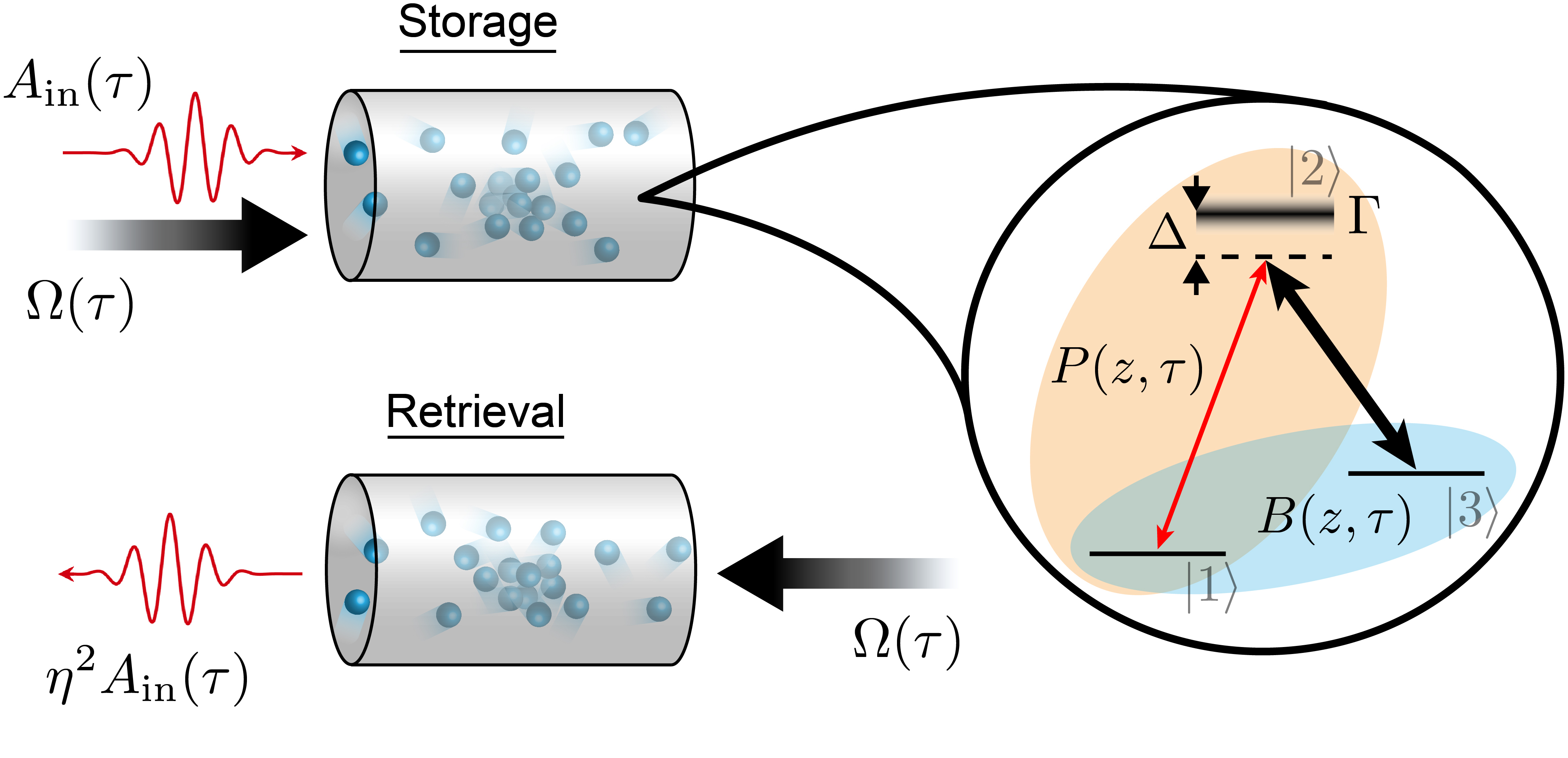}
	\caption{`Backward retrieval' quantum memory scheme, wherein a weak signal field [$A_\text{in}(\tau)$, red thin line] and strong control field [$\Omega(\tau)$, black thick line] enter an atomic medium, generating atomic polarization [$P(z,\tau)$, orange ellipse] and spin wave [$B(z,\tau)$, blue ellipse] fields according to the $\Lambda$-type level scheme shown on the right, with excited-state linewidth $\Gamma$ and two-photon detuning $\Delta$. After a controllable delay, the signal field is retrieved with total efficiency $\eta^2$ via the application of another strong control field that propagates antiparallel to the first control field.}
	\label{fig1}
\end{figure}

In this work, we provide a quantitative performance analysis of resonant $\Lambda$-type quantum memories, shown in Fig.~\ref{fig1}, with a specific focus on signal bandwidths larger than the memory's intermediate state linewidth ($\Gamma$ in Fig.~\ref{fig1}), which we consider broadband. While a variety of other level systems are employed for quantum memory (ladder-type, etc.), $\Lambda$-type level systems are currently the most common, and our analysis is readily generalizable to other level systems. 

In the broadband regime, far off-resonant quantum memory protocols are well-established, but require significantly more control field power than resonant protocols and suffer from low efficiency due to the experimental difficulty in satisfying this requirement \cite{Reim10,England13,Bustard13,England15,Michelberger15,Fisher17,BinBaMemory,Ding15}. In this work, we restrict our discussion to the use of resonant and near-resonant optical fields with Gaussian temporal envelopes in order to avoid the experimental complexities of large pulse energies and shaping of the optical fields. Despite these restrictions, we find that through optimization of the native parameters of Gaussian control fields (i.e., optical power, arrival time, and duration)---which are simple to fine-tune experimentally---we can still achieve high-efficiency memory operation, including in the broadband regime. 

Whereas most work aimed at optimizing quantum memory efficiency focuses on a particular physical quantum memory protocol and leverages physical understanding of the storage mechanism to solve a generic, unconstrained optimization problem \cite{NunnThesis, Gorshkov07_2,Novikova07,Gorshkov07,ATS,ATS2,Vivoli13,BroadbandEIT}, herein we take a physically agnostic approach where we aim to optimize the memory efficiency through a highly constrained set of experimental parameters, initially without regard for the physical storage protocols. While the protocol-based approach facilitates understanding the underlying physics of the quantum memory interaction, practically one is often presented with a set of experimental parameters and resources which are limited, may drift over time, and which in general are not guaranteed to align neatly with a particular storage protocol. Between these physical regimes and storage protocols, it is useful to fine-tune the experimental parameters at hand in order to maximize memory efficiency. 

After numerically calculating the optimal Gaussian control field parameters for a broad range of experimental conditions, we return to examine the physical storage mechanisms and identify the regions of high-efficiency memory operation. We provide physical explanation for the optimized control field parameters in terms of three established memory protocols: those of Refs.~\cite{Moiseev01,Vivoli13,Gorshkov07_2,Carvalho20} that we summarize with the phrase `absorb-then-transfer,' the recently proposed Autler-Townes splitting (ATS) protocol \cite{ATS, ATS2, ATS3, ATS4}, and the electromagnetically induced transparency (EIT) protocol \cite{Fleischhauer02,Phillips01,Lvovsky09,Gorshkov07_2}. As we show, our optimization procedure connects these three physically distinct quantum memory protocols through continuous transformation of the control field parameters (extending the results of Ref.~\cite{ATS2}), and allows for high-efficiency operation in the transition regions between physical protocols. In particular, we report optimized Gaussian control field parameters that allow for optimal memory operation for bandwidths broader than those used in the ATS protocol, and for bandwidths between the ATS and EIT protocols.

This article is organized as follows: After providing details on our numerical analysis of the equations of motion describing the quantum memory interaction and the optimal efficiency bound for a given optical depth (Section \ref{NumericalSec}), in Section \ref{ResultsSec} we consider resonant (Sec.~\ref{resonantSec}) and near-resonant (Sec.~\ref{offresonantSec}) Gaussian control field optimization. In Section \ref{ComparisonSec} we compare the efficiencies generated with the Gaussian optimization described in Sec.~\ref{ResultsSec} and the standard shape-based optimization method described in Refs.~\cite{Novikova07, Gorshkov07, Gorshkov08, Nakao17,NunnThesis}. We find that the Gaussian optimization procedure achieves memory efficiencies comparable to the shape-based method in all but the most broadband cases. In the appendices we provide physical descriptions of the three resonant storage protocols and details on the conditions we use to calculate ATS and EIT regions.

Throughout this work we assume `backward retrieval' of the signal field (Fig.~\ref{fig1}) such that the atomic dynamics during retrieval are the time reverse of those during the storage process, which holds for near-degenerate ground and storage states ($\ket{1}$ and $\ket{3}$ in Fig.~\ref{fig1}) \cite{Gorshkov07, Gorshkov07_2, Gorshkov08}. In this case, the retrieval efficiency is identical to the storage efficiency, $\eta$, and the total memory efficiency is $\eta^2$. Thus, in order to fully characterize the memory efficiency, we need only compute $\eta$. Since the Gaussian fields we consider are intrinsically time-reversal symmetric, under these assumptions no additional experimental measures need to be taken in order to ensure optimization of retrieval beyond routing the retrieval control pulse to the output facet of the atomic ensemble.

\section{Numerical Solution of Maxwell-Bloch Equations}\label{NumericalSec}

The $\Lambda$-type level structure shown in Fig.~\ref{fig1} includes two stable or meta-stable ground states, $\ket{1}$ and $\ket{3}$, and an intermediate excited state $\ket{2}$ that decays to the ground states with the coherence decay rate $\gamma = \Gamma/2$, where $\Gamma$ is the population decay rate of the $\ket{1}$-$\ket{2}$ coherence. All temporal dynamics are considered in the co-moving frame defined by $\tau = t - z/c$, where $t$ is the time measured in the lab frame, $z$ is the one-dimensional spatial coordinate---defined as $z=0$($L$) at the input (output) face of the medium, where $L$ is the medium length---and $c$ is the speed of light. We assume that a control field with frequency near the $\ket{2}\leftrightarrow\ket{3}$ transition, Rabi frequency $\Omega(\tau)$, and duration $\tctrl$ enters the medium with a Gaussian temporal envelope and does not undergo significant absorption or distortion as it propagates [$\Omega(z,\tau) = \Omega(\tau)$]. We assume that before the signal field  enters the medium, it has a Gaussian temporal envelope $A_\text{in}(\tau) = e^{-\tau^2/4\sigma^2}$, where $\sigma=\tFWHM/(2\sqrt{2\ln{2}})$, for the signal duration $\tFWHM$  [temporal full width at half maximum (FWHM)]. There exists also the possibility to temporally chirp the optical field and optimize separately over control pulse bandwidth, but in this work we consider only Fourier-transform-limited pulses such that, e.g., the signal field spectral intensity bandwidth $\delta$ and duration $\tFWHM$ are related by $\delta=2\pi\times 2\ln{2}/(\pi \tFWHM)$. 
 
We further assume that all atoms initially populate the $\ket{1}$ state, which is a valid approximation for atomic populations after optical pumping, or for atomic species with sufficient energy separation between the $\ket{1}$ state and other low-lying states. In general, the signal field undergoes spatial and temporal deformation as it propagates through the medium and is absorbed along the $\ket{1}\rightarrow\ket{2}$ transition, described by $A(z,\tau)$. The atomic dynamics in the presence of these two optical fields are described by the resonant, normalized Maxwell-Bloch equations \cite{Scully,ATS,Gorshkov07,NunnThesis}:

\begin{align}
    \label{Aeq}\partial_z A(z,\tau) &= -\sqrt{d} P(z,\tau)\\
    \label{Peq}\partial_\tau P(z,\tau) = -\bar{\gamma} P(z,\tau) + &\sqrt{d} A(z,\tau) - i\frac{\Omega(\tau)}{2} B(z,\tau)\\
    \label{Beq}\partial_\tau B(z,\tau) = -\gamma_B &B(z,\tau) -i\frac{\Omega^*(\tau)}{2} P(z,\tau),
\end{align}

\noindent where $d$ is the resonant optical depth of the memory, $\bar{\gamma} = (\gamma-i\Delta)/\gamma$ 
is the normalized complex detuning, and $P(z,\tau)$ and $B(z,\tau)$ are macroscopic field operators representing the atomic coherences $\ket{1}\leftrightarrow\ket{2}$ and $\ket{1}\leftrightarrow\ket{3}$, respectively, which are delocalized across the length of the medium. In Eqs.~\eqref{Aeq}-\eqref{Beq}, all frequency (time) scales are normalized by $\gamma$ ($1/\gamma$), and all length scales are normalized by $L$. We assume that the coherence decay rate corresponding to the $\ket{3}\rightarrow\ket{1}$ transition, $\gamma_B$, is negligible during the storage and retrieval operations: $\gamma_B\ll1$.

We iteratively solve these equations of motion using Heun's method for evaluating the $\tau$-derivatives and Chebyshev spectral differentiation for the $z$-derivatives. After integration, we compare the population in $B$ to the population in $A_\text{in}$ in order to calculate the storage efficiency for a particular choice of $\Omega(\tau)$, as:

\begin{equation}
    \label{efficiency}\eta = \frac{\int_0^1 dz\, \abs{B(z,\tau\rightarrow\infty)}^2}{\int_{-\infty}^\infty d\tau\, \abs{A_\text{in}(\tau)}^2},
\end{equation}

\noindent where in practice we truncate $A_\text{in}(\tau)$ and $B(z,\tau)$ at $\tau^\text{end} = 4\tFWHM$, where $A_\text{in}(\tau^\text{end})$ has dropped to $\mathcal{O}(10^{-10})$ of its maximum value. Thus Eqs.~\eqref{Aeq}-\eqref{Beq} in combination with Eq.~\eqref{efficiency} define an objective function that can be maximized with respect to the free parameters of $\Omega(\tau)$. We parameterize the control field Rabi frequency---which we take to be real for simplicity---in terms of its pulse area $\theta = \int_{-\infty}^\infty d\tau\, \Omega(\tau)$, temporal delay $\Deltau$ relative to the arrival of the signal field, and duration $\tctrl=2\sqrt{2\ln{2}}\sigma^\text{ctrl}$
as: 

\begin{equation}
    \label{Omega}
    \Omega(\tau) = \Omega_0\, e^{-[(\tau-\Deltau)/2\sigma^\text{ctrl}]^2},
\end{equation}

\noindent where $\Omega_0 = \theta/(2\sqrt{\pi}\sigma^\text{ctrl})$, and we optimize over the parameter space vector $\mathcal{G} \equiv \left(\theta, \Deltau, \tctrl\right)$ using a Nelder-Mead simplex method, which rapidly identifies the efficiency maxima under these constraints, as verified by deterministic searches of the same parameter space. We define $\tau=0$ at the maximum of the signal field. 
 
Throughout this work we normalize the efficiencies calculated via the method above by the protocol-independent efficiency bound for a fixed optical depth, $\eta_\text{opt}$, described in Refs.~\cite{Gorshkov07_2,Gorshkov07,Phillips08,NunnThesis} and elsewhere. In brief, we calculate this efficiency bound by finding the eigenvalues of the anti-normally ordered storage kernel

\begin{equation}
    \label{etaOptEq}
    K(z,z') = \frac{d}{2}e^{-d(z+z')/2}I_0(d\sqrt{z z'}),
\end{equation}

\noindent where $I_0(x)$ is the zeroth-order modified Bessel function of the first kind, and we discretize $K(z,z')$ on a $5000 \times 5000$ point grid. For fixed $d$, the largest eigenvalue $\lambda_0$ of this kernel represents the maximum achievable storage efficiency at that optical depth, $\eta_\text{opt}=\lambda_0$. By performing this normalization, we aim to compare the efficiencies of particular memory implementations independent of the limitation imposed by finite optical depth.

\section{Results of Gaussian Optimization}\label{ResultsSec}

\subsection{On Resonance ($\Delta = 0$)}\label{resonantSec}

We first consider the case of resonant interaction of the optical fields with the atomic $\Lambda$ system (i.e., $\Delta=0$). At each optical depth and signal bandwidth, we optimize over the control field parameters $\mathcal{G}=\left(\theta,\Deltau,\tctrl\right)$, which fully define any Gaussian control field through Eq.~\eqref{Omega}. This allows us to show that the three known, physically distinct quantum storage protocols for resonant storage (see Appendix \ref{AppendixProtocols} for a brief overview of the protocols) are smoothly connected via continuous transformation of the control-field parameters.  This result is similar to that in Ref.~\cite{ATS2}, which demonstrated ATS and EIT quantum memory behavior can be connected through continuous transformation of the control field Rabi frequency for fixed memory parameters, under the condition of either a constant control field or an interrupted control field of varying linear slope.  Here we distinguish between the \textit{memory} parameters $\mathcal{M}\equiv\left(d,\tFWHM\gamma\right)$, which represent the physical characteristics of a particular quantum memory for the chosen signal bandwidth, and the \textit{control field} parameters $\mathcal{G}$. In this formalism, Ref.~\cite{ATS2} derived a connection between ATS and EIT storage for fixed $\mathcal{M}$ by varying $\mathcal{G}$ [where, e.g., $\mathcal{G}_c=(\Omega_0)$ is a single-parameter vector in the case of a constant control field, $\Omega(\tau)=\Omega_0$]. Motivated by this observation, we consider the distinct condition of Gaussian-shape control fields, and we show that again ATS and EIT memory behavior can be connected if we consider 
the transformation as a function of $\mathcal{M}$, where optimization of $\mathcal{G}$ at each point in $\mathcal{M}$ ensures optimal or near-optimal storage efficiency. Further, we show the two protocols can be connected to the `absorb-then-transfer' protocol through the same continuous transformation. We show each protocol possesses a region of optimality under the restriction of Gaussian pulses and identify two regions where our optimization scheme is most useful: one where the storage mechanism is given by the `absorb-then-transfer' protocol, but in the largely unexplored non-adiabatic regime, and one between the regions of efficient ATS and EIT memory operation.

\begin{figure}[t]
	\centering
	\includegraphics[width=\linewidth]{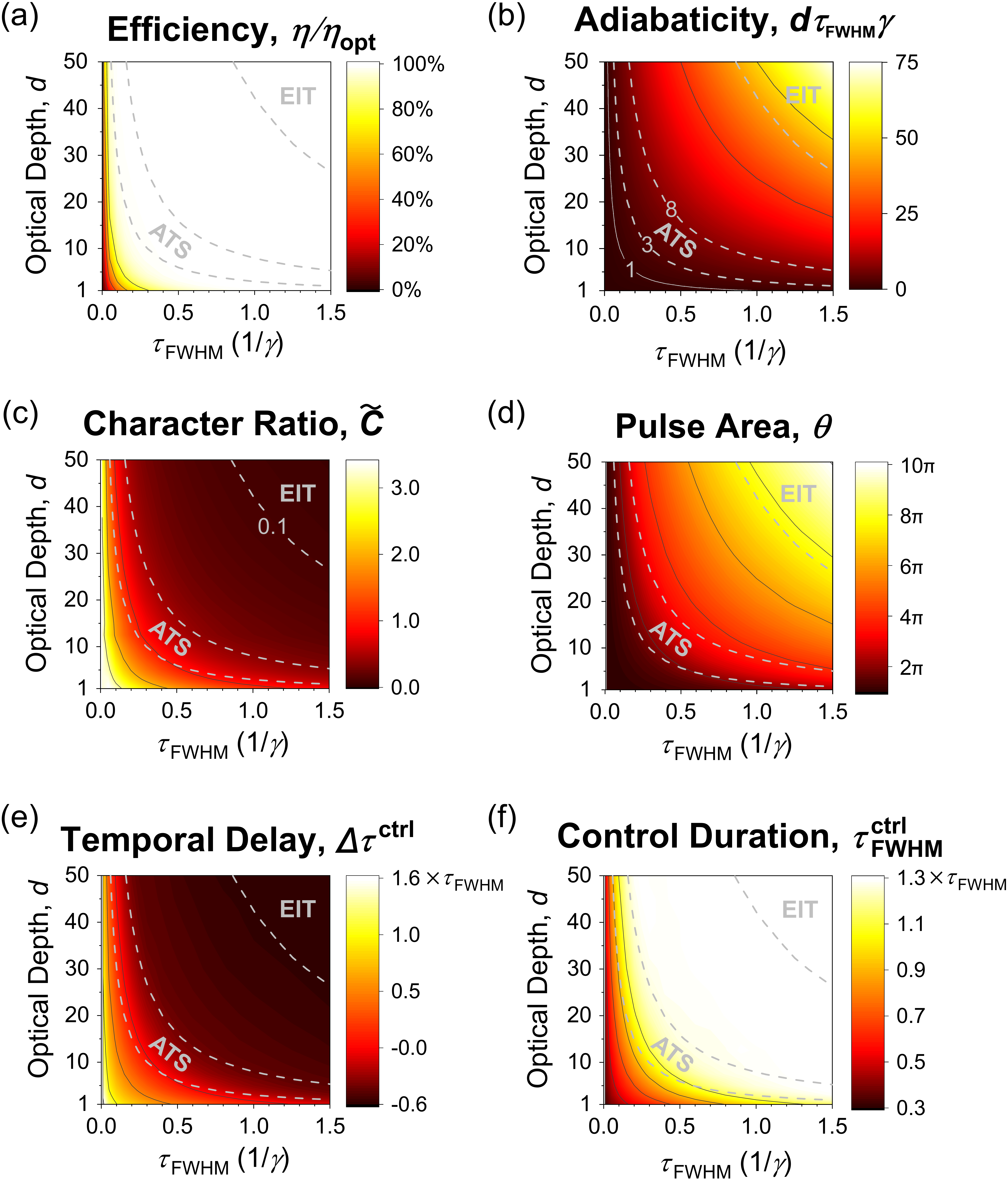}
	\caption{(a) Storage efficiencies achieved through optimization of Gaussian-shape control fields, relative to the optimal bound at each optical depth $d$; (b) memory adiabaticity, where the region $d\tFWHM\gamma=3$ to $8$ corresponds to optimal ATS memory operation (see Appendix B); (c) the memory character ratio, where $\tilde{C}\leq 0.1$ indicates the region of EIT memory operation; and (d)-(f) the optimized control field parameters as a function of optical depth and signal field duration, $\tFWHM$. Positive (negative) delay, $\Deltau>0$ ($\Deltau<0$), refers to control fields that arrive after (before) the signal field. }
	\label{fig2}
\end{figure}

Figure~\ref{fig2} presents the main results of this section. In Fig.~\ref{fig2}(a) we show the normalized efficiencies achieved through the optimization procedure described in Sec.~\ref{NumericalSec}, for memory parameters in the range $d=1$ to $50$ and $\tFWHM\gamma=0$ to $1.5$, which we take to be representative of the bulk of experimental broadband quantum memories, though our analysis is easily extended to other regions. The efficiencies shown saturate the optimal efficiency bound ($\eta/\eta_\text{opt}=100\%$) over the whole memory parameter space we investigate, as long as the adiabaticity criterion $d\tFWHM\gamma\geq1$ is satisfied. This result demonstrates that Gaussian-shape fields are sufficient for high-efficiency, broadband memory operation, without the need for full pulse-shape control.

The first region of the $\mathcal{M}$ parameter space where our optimization is most useful can be highlighted using Fig.~\ref{fig2}(b), which shows the memory adiabaticity ($d\tFWHM\gamma$) as a function of $\mathcal{M}$. For $d\tFWHM\gamma<1$, we observe the expected decay of the storage efficiency \cite{Gorshkov07_2}, shown in Fig.~\ref{fig2}(a). Between $d\tFWHM\gamma=1$ and the region of efficient ATS operation [delineated with dashed lines in Fig.~\ref{fig2}(b)---see Appendix \ref{AppendixATSRegion} for derivation], we observe storage efficiencies that approach the optimal bound ($\eta/\eta_{opt} = 100\%$), where the physical storage mechanism is given by the `absorb-then-transfer' protocol \cite{Moiseev01,Vivoli13,Gorshkov07_2,Carvalho20}. As can be seen in Fig.~\ref{fig2}(d)-(f), the optimized control field parameters in this region correspond to approximately $\pi$-pulse-area control fields that are narrower in duration than the signal fields they store ($\tctrl<\tFWHM$), and arrive after the signal field ($\Deltau>0$). The optimized control fields arrive at the approximate time when the electric field of the signal changes sign (when evaluated at $z=1/2$), in agreement with the analysis of Refs.~\cite{Carvalho20,ZeroAreaPhotons,ZeroAreaPhotons2}. This result demonstrates that the absorb-then-transfer protocol can approach the optimal efficiency bound in the non-adiabatic regime, in addition to the adiabatic regime investigated in Ref.~\cite{Vivoli13}. 

The second region of $\mathcal{M}$-space where our optimization procedure is most useful is in the region between the optimal memory conditions for ATS and EIT storage, delineated by the dashed lines in Fig.~\ref{fig2} (see Appendix ~\ref{AppendixEITRegion} for derivation of the boundary of the EIT region). Here the memory is still non-adiabatic ($d\tFWHM\gamma\gg1$ is not satisfied), but the ATS condition $d\tFWHM\gamma = 3$ to $8$ is exceeded, similar to the broadband-EIT region of Refs.~\cite{ATS2,BroadbandEIT}. In this region, fine-tuning of the control field parameters allows for optimal memory efficiency, whereas use of the typical ATS [$\mathcal{G} = (2\pi,0,\tFWHM)$] or EIT control field parameters leads to sub-optimal efficiency.

We note that Fig.~\ref{fig2}(d)-(f) may act as a guide for experimentally simple optimization of broadband quantum memory using Gaussian pulses. For a given set of memory parameters $\mathcal{M}$, the optimal Gaussian control field parameters may be read off directly from Fig.~\ref{fig2}(d)-(f). In the adiabatic, EIT regime ($d\tFWHM\gamma \gg 1$), we find negative temporal delays that asymptote to around $-0.55\tFWHM$, and control field durations that asymptote to $\sim1.33\tFWHM$. 

Notably, the optimized control field parameters presented in Fig.~\ref{fig2} are not mutually independent; for example, at fixed $\mathcal{M}$, changes to the control field duration away from the optimal choice shown in Fig.~\ref{fig2}(f) may be compensated for with changes to the control field delay and pulse area, with only a small decrease in efficiency in some cases. Fig.~\ref{fig2}(d)-(f) shows only the optimal and mutually dependent choice of control field parameters. The sensitivity or robustness to noise in these optimal parameters may be the subject of future work. 

\subsection{Near Resonance ($\Delta \neq 0$)}\label{offresonantSec}

\begin{figure}[t]
	\centering
	\includegraphics[width=\linewidth]{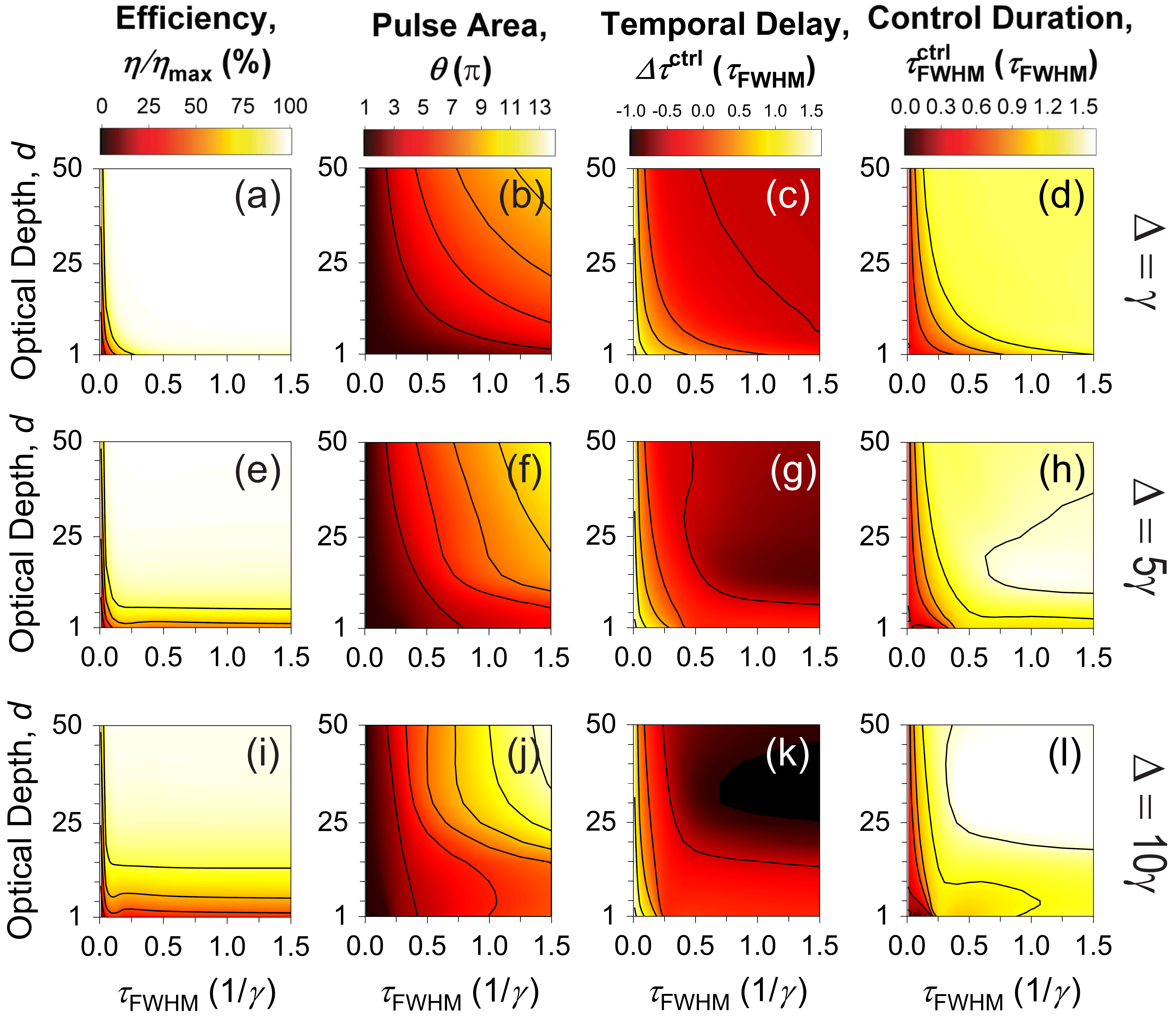}
	\caption{Optimized $\Lambda$-type quantum memory efficiency and corresponding optimized control field parameters as a function of optical depth, $d$, and signal field duration, $\tFWHM$, for varying two-photon detuning, (a)-(d) $\Delta=\gamma$, (e)-(h) $\Delta=5\gamma$, and (i)-(l) $\Delta=10\gamma$, for excited state coherence decay rate $\gamma$.}
	\label{fig3}
\end{figure}

The case of resonant signal and control fields considered above has shown optimal storage efficiency to be possible for a wide range of memory parameters using only Gaussian pulses. In this section we continue this analysis for non-zero two-photon detunings in the near-resonant regime, where $\Delta$ is of order $\gamma$. This analysis differs then from the far-off-resonant Raman regime \cite{NunnThesis,England13,Bustard13,England15,Michelberger15,Fisher17}, where $\Delta\gg\gamma$. While we nominally only consider positive detunings, the results presented in this section are symmetric about $\Delta=0$.

Whereas for resonant signal and control fields all optimization parameters $\mathcal{G}$ are smooth, monotonic functions of the memory parameters $\mathcal{M}$, in the near-resonant case we observe more complicated behavior where the optimized parameters are no longer strictly monotonic functions of $\mathcal{M}$. In Fig.~\ref{fig3}(a)-(d), (e)-(h), and (i)-(l), we consider two-photon detunings $\Delta=\gamma$, $\Delta=5\gamma$, and $\Delta=10\gamma$, respectively. As shown in Fig.~\ref{fig3}(a), (e), and (i), for fixed memory parameters $\mathcal{M}$, larger $\Delta$ consistently implies smaller $\eta/\eta_\text{max}$. It appears this decrease in memory efficiency can be avoided by increasing optical depth, although this comes at the cost of larger control field pulse area and optical power required to implement optimized storage. In general, as shown in Fig.~\ref{fig3}(b), (f), and (j), for fixed $\mathcal{M}$ the pulse areas necessary to implement optimized quantum storage with Gaussian pulses tend to increase with increasing $\Delta$. The minimum temporal delay in the region we simulate decreases as a function of $\Delta$, indicating control field pulses in those regions of negative delay that arrive significantly sooner (before the signal field) than their resonant counterparts in the EIT regime. In these same regions [i.e., $\mathcal{M}\approx(20, 1.5)$ for $\Delta=5\gamma$, $\mathcal{M}\approx(35, 1.5)$ for $\Delta=10\gamma$], the control field duration is also significantly larger than in the resonant case.  

As in Sec.~\ref{resonantSec}, we note that Fig.~\ref{fig3} may serve as an experimental guide for optimized quantum memory implementation with Gaussian-shape signal and control fields at fixed detuning in the near-resonant regime.

\section{Comparison of Gaussian and Shape-Based Optimization}\label{ComparisonSec}

In the sections above we have introduced an alternative quantum memory optimization scheme that relies only on broadband light pulses with Gaussian temporal envelope. This scheme operates at or near two-photon resonance, and therefore avoids the experimental complexities associated with full pulse-shape control of broadband fields and the use of large pulse energies. In this section, we compare the results of this optimization scheme with the more standard shape-based optimization of Refs.~\cite{Novikova07, Gorshkov07, Gorshkov08, Nakao17,NunnThesis}. 

In order to enumerate this comparison, we consider a quantum memory with optical depth $d=50$, where we calculate via Eq.~\eqref{etaOptEq} an optimal storage efficiency of $\eta_\text{opt}=95.2$\% (total efficiency: $\eta_\text{opt}^2 = 90.6$\%). We further assume resonant storage of photons such that $\Delta=0$. To calculate the storage efficiencies achieved via shape-based optimization, we first numerically construct the storage kernel $K(z,\tau)$ defined by the linear integral transform,

\begin{equation}
   B_\text{out}(z)=B(z,\tau\rightarrow\infty) = \int_{-\infty}^{\infty} d\tau\, K(z,\tau) A_\text{in}(\tau),
\end{equation}

\noindent via the method described in Ref.~\cite{NunnThesis}. The largest singular value of $K(z,\tau)$ and the corresponding right-singular vector represent the optimal storage efficiency and optimal signal mode temporal profile, respectively \cite{NunnThesis,Gorshkov07_2}. Importantly, $K(z,\tau)$ depends both on the chosen optical depth, $d$, and the control field parameters, $\mathcal{G}$. The optimal signal mode calculated through this method is therefore guaranteed to lead to optimal storage efficiency for the given $d$ and $\mathcal{G}$.

This method relies on signal-field shaping in order to achieve optimal memory efficiency. One can instead optimize the memory efficiency through shaping of the control field with the procedure outlined in Ref.~\cite{NunnThesis}. In short, one interpolates between the optimal signal mode calculated for given $d$ and $\mathcal{G}$ and the desired signal mode (typically a Gaussian, with duration $\tFWHM$), and at each interpolation step one optimizes $\mathcal{G}_s$, which is a large vector that defines the shape of $\Omega(\tau)$. At each interpolation step, the signal field is deformed away from the optimal shape and the control field shape is optimized in order to compensate for the decrease in memory efficiency. For sufficiently small successive deformations of the signal field, optimality is preserved at each interpolation step and this procedure leads to the optimal control field shape for a Gaussian signal field. We find the final memory efficiency achieved through control-field shaping is typically bounded above by the efficiency achieved through signal-field shaping. For the purposes of this comparison, we compare the results of the Gaussian optimization in Sec.~\ref{ResultsSec} with the upper bound achieved via signal-field shaping.

\begin{figure}[t]
	\centering
	\includegraphics[width=0.9\linewidth]{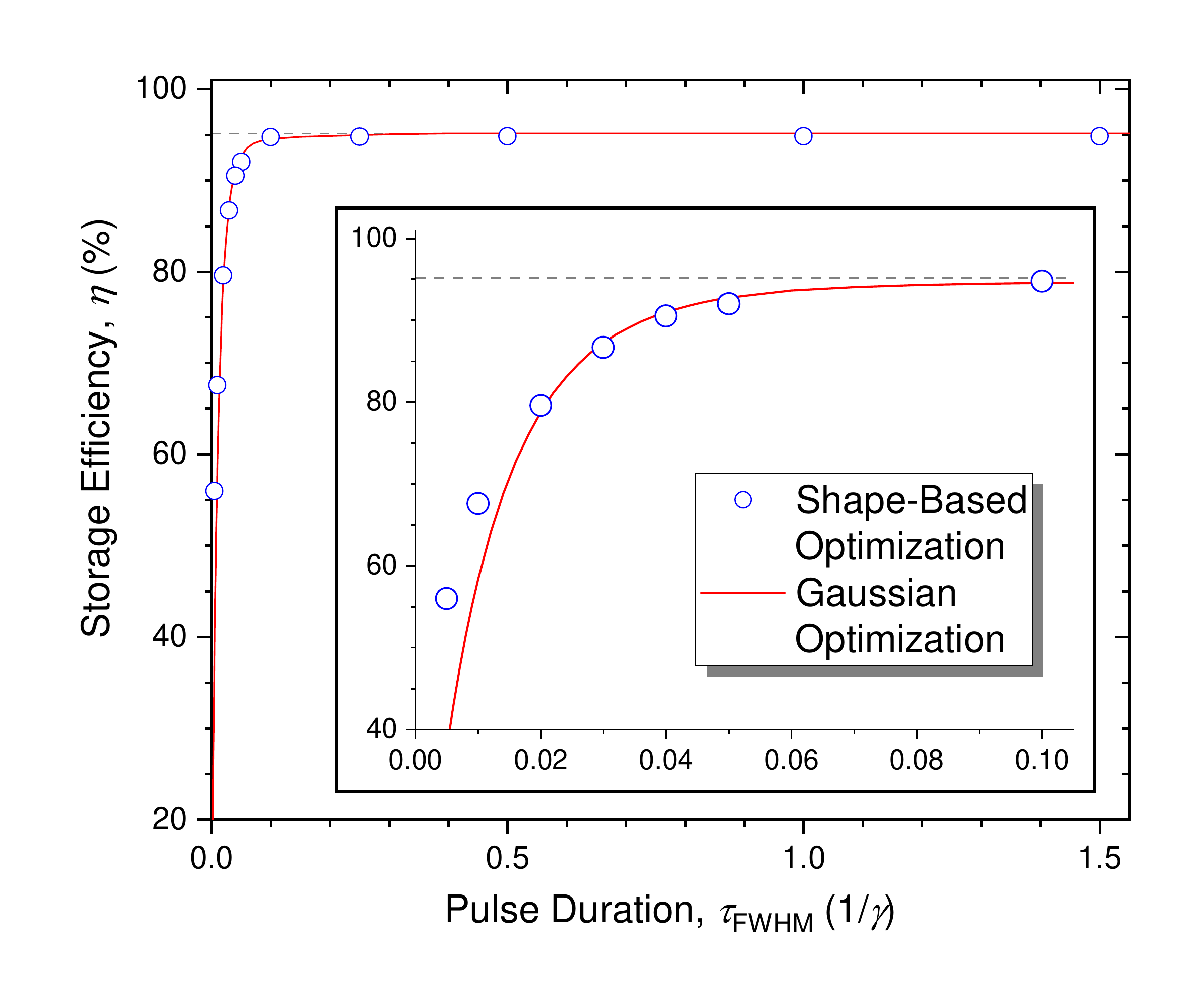}
	\caption{Comparison of quantum memory efficiencies achieved using shape-based and Gaussian control field optimization for d=50. Dotted line represents the optimal efficiency bound, $\eta_\text{opt}=95.2$\%.}
	\label{fig4}
\end{figure}

In Fig.~\ref{fig4}, we calculate the storage efficiency achieved via signal-field shaping alongside the efficiency calculated via the Gaussian optimization scheme presented in this article, for signal durations between 0 and $1.5/\gamma$ in the example case of a $\Lambda$-type level system at $d=50$. We observe saturation of the optimal bound (dashed line, $\eta=\eta_\text{opt}=95.2\%$) in the region $\tFWHM = 0.1/\gamma$ to $1.5/\gamma$ for both optimization schemes, where the optimal storage protocol transitions between all three resonant protocols defined in Appendix~\ref{AppendixProtocols}. Below $0.1/\gamma$ signal field duration, we observe decay in the memory efficiency for both schemes, where the Gaussian optimization scheme leads to comparatively lower storage efficiencies for the most broadband pulse durations. Nevertheless, the Gaussian optimization procedure provides comparable memory performance over a wide range of bandwidths.

The main result of this section is as follows: Through the Gaussian optimization procedure described in this article, we achieve storage efficiencies that closely compare with the efficiencies achieved through shape-based optimization, but which (1) require significantly less computational expense to calculate, and (2) physically correspond to quantum memory experiments that are simpler, as they eliminate the need for arbitrary shaping of either intense, broadband fields (control-field shaping) or broadband single-photon level fields (signal-field shaping).

\section{Conclusions and Future Work}

In this work we have presented a quantitative and qualitative exploration of $\Lambda$-type quantum memory with Gaussian optical fields that are resonant and near-resonant with an atomic two-photon transition. The restriction to Gaussian fields serves to simplify experimental implementations of optical quantum memory. We have shown that despite this restriction, optimization of the parameters of Gaussian control fields (optical power, arrival time, and duration) can lead to high-efficiency memory operation over a wide range of memory parameters in the broadband regime. We make the distinction between the memory parameters ($\mathcal{M}$) and the control field parameters ($\mathcal{G}$), and in so doing we find that optimization of $\mathcal{G}$ reveals a continuous transition between three physically distinct quantum memory protocols (what we call the `absorb-then-transfer' protocol, ATS, and EIT) as a function of $\mathcal{M}$. This optimization procedure is most useful in two regions of the memory parameter space. In the region of $\mathcal{M}$ that is more broadband than the optimal ATS region at fixed optical depth, we show that the `absorb-then-transfer' protocol can operate with near-optimal efficiency, extending the result investigated previously in the adiabatic ($d\tFWHM\gamma\gg1$) regime \cite{Vivoli13}. In the region of $\mathcal{M}$ between optimal ATS and EIT operation---the mixed ATS/EIT regime---we also show that optimal memory operation is possible.

In Sec.~\ref{offresonantSec} we have extended this analysis to the near-resonant regime where the two-photon detuning is of order the excited state linewidth, $\Delta\sim\gamma$. We observe similar qualitative behaviour of the optimal control field parameters $\mathcal{G}$ as a function of $\mathcal{M}$, but in order to achieve the same memory efficiency, a larger optical depth $d$ and control field pulse area $\theta$ are required compared to the resonant case. 

Finally, in Sec.~\ref{ComparisonSec} we have provided a numerical comparison of the proposed Gaussian optimization technique with the more common shape-based optimization procedure. We find that Gaussian pulses are suitable for optimal memory operation over a wide range of memory parameters, and only perform significantly worse than arbitrarily shaped pulses in the most broadband cases where the storage efficiency is non-optimal even for shape-based optimization. 

In this work we restrict ourselves to the widely available resource of Fourier-transform limited pulses, where pulse duration and bandwidth are Fourier-transform pairs and accordingly only describe one degree of freedom subject to optimization. Future work may consider optimization via chirped optical fields, which expands the toolbox for optimization of Gaussian quantum memory and has been explored in other memory protocols \cite{Minar10, Demeter14,Zhang14}. We have also restricted our optimization procedure to the case of homogeneous dephasing of the atomic polarization field. The case of inhomogeneous polarization dephasing, following the approach of \cite{Gorshkov07_3}, may be considered in future work.

\section*{Acknowledgements}

We gratefully acknowledge helpful discussion provided by Yujie Zhang, Xinan Chen, Sehyun Park, Elizabeth Goldschmidt, Bin Fang, Shuai Dong, Seth Meiselman, and Offir Cohen, as well as support from NSF Grant Nos. 1640968, 1806572, 1839177, and 1936321 and NSF Award DMR-1747426.

\vspace{1em}
\appendix

\renewcommand\thefigure{\thesection.\arabic{figure}}

\section{Description of Protocols}\label{AppendixProtocols}
\setcounter{figure}{0} 

Here we briefly review the three known resonant quantum memory protocols that make use of a homogeneously broadened excited-state linewidth, and their key features:

\textit{(1) `Absorb-then-transfer.'} Described in Refs.~\cite{Moiseev01,Vivoli13,Gorshkov07_2,Carvalho20}, quantum storage is achieved through linear absorption of the signal field along the $\ket{1}\rightarrow\ket{2}$ transition and coherent population transfer between the atomic polarization and spin-wave field via a $\pi$-pulse control field. We distinguish this `absorb-then-transfer' storage protocol from the related photon-echo protocols \cite{spinecho1,spinecho2,spinecho3,AFC1}, as photon emission upon retrieval does not depend on dipole rephasing for homogeneously broadened intermediate states. Ref.~\cite{Carvalho20} indicates that in order to optimize storage efficiency the arrival time of the control field should occur near the first zero of the complex signal field amplitude when evaluated at the middle of the ensemble ($z=1/2$), at least in the weak-absorption regime. Ref.~\cite{Vivoli13} has shown this storage protocol can be optimal (i.e., can achieve $\eta=\eta_\text{opt}$) for large optical depths, such that $d\tFWHM\gamma\gg1$.

\textit{(2) Autler-Townes Splitting.} In the recently proposed Autler-Townes-Splitting (ATS) protocol \cite{ATS,ATS2,ATS3,ATS4}, a control field propagates with the signal field at zero delay ($\Deltau=0$) with pulse area $\theta = 2\pi$, creating an Autler-Townes doublet in the signal field absorption profile that matches the signal field bandwidth. As shown in Ref.~\cite{ATS2}, one is free to choose any control-field shape, as long as $\theta = 2\pi$ is fulfilled over the duration of the signal field. As more broadband signal fields experience lesser effective optical depth (due to increasing Autler-Townes splitting), and more narrowband pulses lead to decoherence of the atomic polarization during the storage operation, the ATS protocol is constrained to optimal operation in a narrow bandwidth region around a unique choice of $\tFWHM$ for a given optical depth \cite{ATS,ATS2} (see Appendix B).

\textit{(3) Electromagnetically Induced Transparency.} The well-known Electromagnetically Induced Transparency (EIT) protocol is described in the narrowband regime in Refs.~\cite{Fleischhauer02,Phillips01,Lvovsky09,Gorshkov07_2}, and in the broadband regime in Refs.~\cite{ATS2,BroadbandEIT}. A control field of duration longer than the signal field ($\tctrl>\tFWHM$) enters the medium ahead of the signal field in time ($\Deltau<0$) and opens a spectral transparency window that is slowly closed after the signal field enters the medium, thereby trapping the signal field in the medium via the slow-light effect.

We note that the key physical features of all three protocols are compatible with the use of Gaussian-shape control fields, which helps to explain why Gaussian-shape fields are sufficient to achieve the high storage efficiencies of Sec. \ref{ResultsSec} and \ref{ComparisonSec}.

\section{Derivation of the ATS Region}\label{AppendixATSRegion}
\setcounter{figure}{0} 

\begin{figure*}[t]
	\centering
	\includegraphics[width=0.75\linewidth]{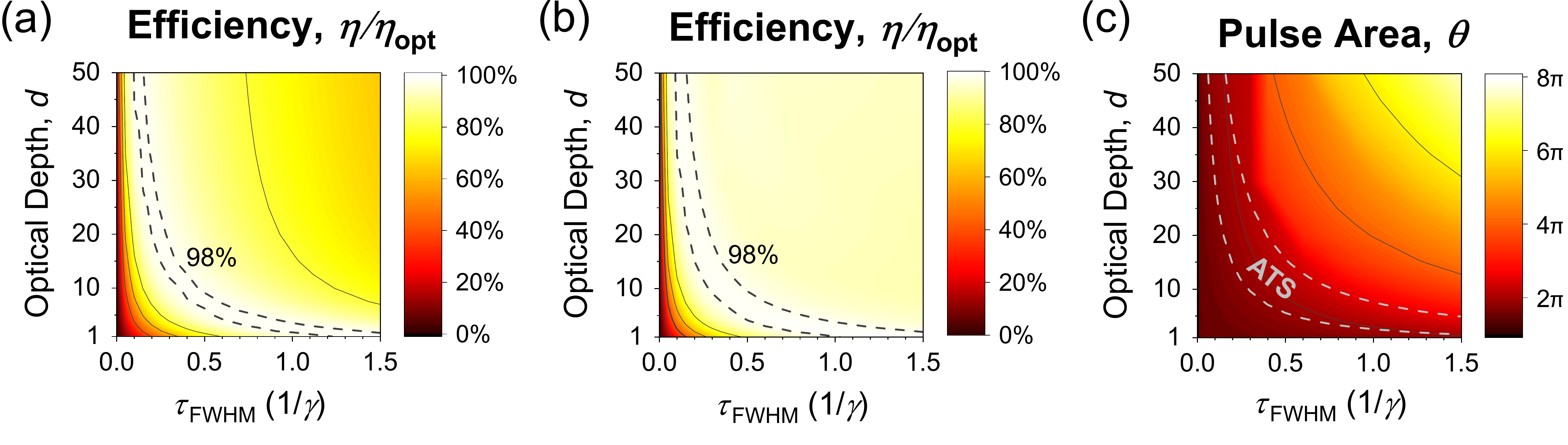}
	\caption{Comparison of quantum memory efficiencies achieved using pulse areas of (a) $\theta=2\pi$, and (b) directly optimized pulse areas shown in (c). Dotted lines in (c) represent the ATS condition $d\tFWHM\gamma=3$ to 8 used in the main text, which captures the region of $\theta\approx 2\pi$ operation.} 
	\label{appendixFig1}
\end{figure*}

As stated in Appendix A, at fixed optical depth the ATS protocol is constrained to optimal operation in a narrow region around a unique value of $\tFWHM$ \cite{ATS,ATS2}. In this appendix we derive an approximation of this region, for the memory parameters investigated in the text, given by $d\tFWHM\gamma = 3$ to 8. Notably, this approximation is dependent on the region of $\mathcal{M}$ chosen, and is not strictly valid for other regions, such as those in Refs.~\cite{ATS,ATS2,ATS3}.

Nominally the ATS protocol requires pulse areas $\theta=2\pi$ for optimal operation, however, as stated in Ref.~\cite{ATS}, this constraint is relaxed in regions of non-optimal effective optical depth $\tilde{d}<3$, where $\tilde{d} = d\tFWHM\gamma \, \pi/ (2 \theta \ln2 )$ for Gaussian pulses. If the effective optical depth is small---as is frequently the case in the broadband regime considered here---pulses with area $\theta<2\pi$ yield larger memory efficiency than $\theta=2\pi$, due to a reduction in Autler-Townes splitting and an increase in the effective optical depth. 

In order to accurately capture the ATS region of $\mathcal{M}$ discussed in Section \ref{resonantSec}, we do not rely solely on the region of high-efficiency operation with $\theta=2\pi$ control fields. Instead, we follow a reduced version of the optimization procedure presented in the main text, wherein we fix $\Deltau=0$ and $\tctrl=\tFWHM$ and optimize over $\theta$. The efficiencies resulting from this optimization procedure, compared to the efficiencies for $\theta=2\pi$ only, are presented in Fig.~\ref{appendixFig1}(a)-(b). For purposes of comparison, we plot the boundary of $\eta/\eta_\text{opt}=98\%$ operation, which is increased upon optimization of $\theta$. The optimized pulse areas corresponding to Fig.~\ref{appendixFig1}(b) are shown in Fig.~\ref{appendixFig1}(c). The region of $\theta\approx2\pi$ operation which results in high normalized efficiency is well-captured by the condition $d\tFWHM\gamma=3$ to $8$ in this region of $\mathcal{M}$, and accordingly we take this condition to be representative of ATS operation for the memory parameters under consideration.

We note that for extremely low optical depths, $d\sim1$, the region of high-efficiency ATS operation diverges from the $d\tFWHM\gamma=3$ to $8$ condition towards smaller adiabaticity. Other conditions may be used to define the ATS region, for example based on optimal delay (e.g., $\Deltau=-0.25$ to $0.25$) or the character ratio discussed in Appendix~\ref{AppendixEITRegion} (e.g., $\tilde{C} = 0.75$ to $1.25$), that better capture this region of ATS operation. However, the region given by $d\tFWHM\gamma=3$ to $8$ is the largest of these 
and is consistent with Refs.~\cite{ATS,ATS2,ATS3}.

\section{Derivation of the EIT region}\label{AppendixEITRegion}
\setcounter{figure}{0} 

We define the boundary of the EIT region via the character ratio

\begin{equation}
C=\frac{1}{\tau_s}\frac{\int_{-\tau_s/2}^{\tau_s/2}d\tau \int_0^{1}dz\,\abs{P(z,\tau)}^2}{ \int_0^{1}dz\,\abs{B(z,\tau\rightarrow\infty)}^2}
\end{equation}

\noindent introduced in Ref.~\cite{ATS2}, which gives the ratio of the transient population that enters $P(z,\tau)$ during the storage period $\tau_s = 2.25 \tFWHM$ to the population that arrives in $B(z,\tau)$ after the storage operation is completed. We consider the normalized character ratio $\tilde{C}={C}/{C}_0$ using the value of $C$ for each optical depth that corresponds to `pure' ATS operation with $\Deltau=0$, which we identify as ${C}_0$. 
Using this normalization, we consider the region of $\mathcal{M}$ where $\tilde{C}\leq0.1$ to correspond to EIT operation, shown in the shaded region of Fig.~\ref{fig2}(c) \cite{ATS2}. In this region, Fig.~\ref{fig2}(d)-(f) shows the optimal control fields have larger pulse area than in the `absorb-then-transfer' or ATS regions, the control fields are broader in duration than the accompanying signal field, and the control fields arrive before the signal field. This behavior is a signature of EIT storage, and supports the choice of $\tilde{C}\leq0.1$ as the threshold for EIT behavior.


\providecommand{\noopsort}[1]{}\providecommand{\singleletter}[1]{#1}%

\end{document}